# Analysis of challenges faced by WebRTC videoconferencing and a remedial architecture

Maruf Pasha[1], Furrakh Shahzad[2], Arslan Ahmad[3]

([1]Department of Information Technology, Bahauddin Zakariya University, Multan 60000, Pakistan)

([2]Department of Computer Science, Pakistan Institute of Engineering and Technology, Multan 60000, Pakistan)

maruf.pasha@bzu.edu.pk, farrukh.khan0913@gmail.com, arslan_ahmad91@yahoo.com

***Abstract*----Lately, World Wide Web came up with an evolution in the niche of videoconference applications. Latest technologies give browsers a capacity to initiate real-time communications. WebRTC is one of the free and open source projects that aim at providing the users freedom to enjoy real-time communications, and it does so by following and redefining the standards. However, WebRTC is still a new project and it lacks some high-end videoconferencing features such as media mixing, recording of a session and different network conditions adaptation. This paper is an attempt at analyzing the shortcomings and challenges faced by WebRTC and proposing a Multipoint Control Unit or traditional communications entity based architecture as a solution.**

## 1. INTRODUCTION

Videoconferencing (VC) is another term used for conducting a videoconference or video teleconference with the help of a set of telecommunication equipment and technologies which facilitate two or more places to communicate with each other with the help of simultaneous and mutual audio video transmissions [3,6]. It is also termed as 'visual collaboration' and sometimes it is also referred as one of the kinds of groupware [25]. Throughout the last ten years, we have witnessed the evolution of World Wide Web. In the past web pages were simply supposed to be a piece of static information but now they have become top-notch applications with unprecedented functionality and interactivity.

This trend obviously also affected the videoconference. Now out of nowhere multimedia communication windows open in the browsers with full support of proprietary plugins e.g. Adobe Flash [1]. On one hand these products provide full-fledged videoconferencing and collaborating exposure, while on the other hand they frequently depend on the solutions such as Adobe Flash and the protocols forced by the browser plugins developers, instead of the familiar standards prevalent in the industry. Recently, there have been enough efforts done which gives web pages and browsers more interactivity and functionality without relying on the plugins.

HTML5 attempts at providing web pages with advanced features by making improvements in the typical markup language and adding the APIs of JavaScript. Hence, there have been efforts of offering some real-time and mutual communication between peers, to the WWW. WebRTC is developed and defined to do this job by using the functionalities of HTML5, other existent technology, real-time procedures and codes despite of reinventing the wheel: forming the novel protocols and codes [21, 7]. WebRTC comes up as one of the combined efforts initiated from the WebRTC of WWW Consortium (W3C) and also from rtcweb that belongs to IETF (Internet Engineering Task Force).

In this setting, the former provides with JavaScript API and HTML5 while the latter outlines the codes and protocols that are going to be used in the communication [2]. Nowadays, the first blueprints or applications of WebRTC are under development. However, it is safe to say that the outlines of this new communication setup are not in a complete form and there is still a lot of room for creativity in respect of unconventional communication services.

This is where this paper comes into the picture. This paper is our meek effort to offer a synopsis of the videoconference through WebRTC when compared to the traditional real-time communication practices within the web browsers. Moreover, we enlist the trials, problems and challenges faced by this new technology, its users, and we came up with a perfect solution for the said challenges through re-introducing the very well-known communication technology named as MCU which is short for Multiple Control Unit [24, 20]. Lastly, we devise architecture for the MCU and present the possible first steps that may take us towards implementing this striking architecture.

## 2. WEB VIDEO CONFERENCE

RIAs (Rich Internet Applications) [11] are often taken as the dawn of static pages. These procedures and applications makes our jobs easy by providing us with text simulations, input treatment by user, drawings, drag & drop functions and a bi-directional flow plus streaming of the audio-video content amongst all remaining features. Microsoft Silverlight, Oracle JavaFX and Adobe Flash are the current top famous platforms with more than 50% of market penetration. Unlike these proprietary solutions, W3C works hard on shaping the next form of HTML markup language that is HTML5 which promises to offer more desktop type features in a normal and acceptable fashion to web pages.

WebRTC is the joint venture of ERTF and W3C which promises to offer the browsers with a peer to peer and real-time communicating experience. They target to specify the standard quantity of procedures, protocols (to be done by IETF) and JavaScript APIs (to be done by W3C) that present some real-time communication abilities for World Wide Web developers to use in their certain applications. Contrary to the traditional alternatives, WebRTC is actually defined as a system that uses prevalent, industry leading, famous and successfully tested standards: not the exclusive solutions. WebRTC with and without media server is shown in Fig. 1.

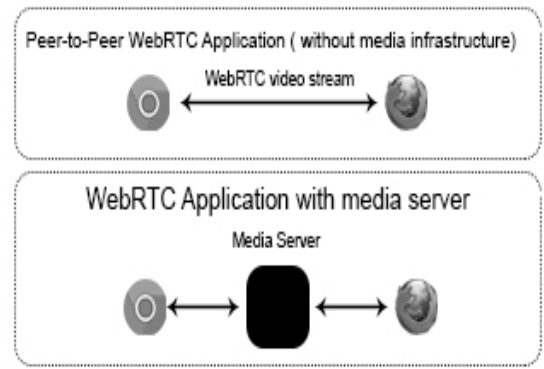

Fig. 1 WebRTC Interactions

Some of the very important and applicable decisions that were taken while jotting down this paper are as follow:

- ICE [14] along with TURN and STUN is approved to be used for NAT (Network Address Translation) Traversal.
- RTP [8] and the safe variation SRTP [17, 9] will be tried to frame the data.
- Media negotiation is going to be held through SDP [18]
- Not a single signaling procedure or protocol (e.g. SIP [13, 16]) is recommended so that the developers enjoy more elasticity in creativity for their web-based operations and applications.

Although the procedure of defining this setup is not complete, the executions are already being presented and the video conferencing is now reached at first milestone. In this paper, we are going to overview the basic details of WebRTC in respect of videoconferencing and challenges it faces. We are also going to propose the solutions for those challenges.

## 3. CHALLENGES FACED BY VIDEOCONFERENCING

When one goes about applying one of the recognized technologies within a totally different ecosystem, it always brings a lot of problems and challenges. The past experiences [12] and [22] give us optimum insight and information to predict the problems and challenges that should be addressed before videoconferencing via WebRTC becomes broadly used in any cutting-edge applications which increase the level of one-to-one communications. The challenges that we are going to discuss, belong to entire web ecosystem where preferably various types of the equipment and device efficiently interact and provide high level services which might be needed in complicated scenarios e.g. education:

3.1. Diverse Access Devices and Networks

With the passage of time, the diversity of devices that access the web has reasonably increased. There was a time when people could access WWW though personal computers only,, but now there is a wide range of devices with many different capacities and access networks. For example, consider smart phones that have only a limited process power as compared to desktops or laptops, and yet they access the WWW via certain cellular networks.

3.2. Screen Size Issues

When it comes to screen resolutions and quality of the images, smaller screens would not show the exact amount of information compared to big screens. Same goes for videoconferencing, sending a high quality video to a person who uses a small screen, would be simply useless since the user will not be able to feel the difference. In complicated scenarios, we can take multi-conferencing as an example; it is not possible to show all the participants at the same time using small screen.

3.3. CPU Problems

WebRTC videoconferencing is a real-time communication; it demands the processor to do decoding, encoding and dispensing of the video and audio at the same period of time. We define this as CPU stress and it relies on many factors e.g. used codec's, quality of the video, audio and their respective sizes. The problem gets worse in case of cell phones; not only the cell phone CPUs impose a limitation at how the processes can be carried out, but also a battery drain due to CPU stress.

3.4. Availability of Bandwidth and Latency

As mentioned above, a wide array of devices are used with varying networks to access. As an example, we use wired Ethernet connection for desktops computers and 3G for cell phones. These differences might not seem important, but they must be considered if communication optimization is our goal. To make things interesting, let us consider a 3G networks; their currently available bandwidth without any notice may vary from previously available bandwidth.

4. IMPLEMENTATION OF MCU

To address the above-mentioned challenges, we are trying to work out the implementation of MCU (Multipoint Control Unit) capable of adapting, redirecting and translating streams of media. While this paper is being written, WebRTC IETF group has already defined almost all of the features of the videoconferencing at the level of protocol and media negotiation. Likewise, the implementation of MCU has to be one of the safe routes since not only it works at the transport layer but also the layers underneath it. Hence leaving flexibility to the level of application instead of relying on the development of the definition of JavaScript API offered by the W3C.

What we suggest hereby is making of a novel Multipoint Control Unit which adheres to the standards, and is capable to interconnect ecosystem of WebRTC as an enabler that provides the out of the box services. Browsers interaction with each other for video conferencing by using the MCU media server is shown is Fig. 2.

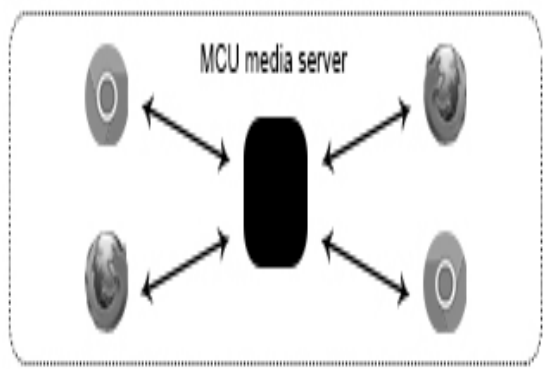

Fig. 2 Browsers interaction with MCU

### 4.1. Mechanism of MCU

The Multiple Control Unit is a very important part of multipoint videoconference from the day it was conceived. In simple words it is the central device which weaves the interconnection among the various members taking part in videoconferencing as depicted in [19]. It becomes a prerequisite in the case where the clients can establish none but only a connection based on single-point. In the perspective of this research work, Multiple Control Unit is actually a piece of advanced software which, by ensuing the explanations of IETF, uninterruptedly communicates with WebRTC while in doing so it is able to redirect and transcode the media streams.

Multiple Communication Unit is likely to be found at any place within the network, and as doesn't represent any of the participants of a conference; it is actually nothing but an entity that performs all the tasks just to provide some cutting-edge communication services. Even if we discuss the possibility of WebRTC peers making some network of the overlay type by establishing two or more connections, the part of a Multiple Communication Unit still make some sense and obviously it is very important when it comes to tackling the challenges as shown above.

In the easiest to understand scenario, a Multiple Communication Unit has to redirect the media flows among the people taking part in communication, by only enabling them to gather the data from other participants, acting as the middle point of a matrix. As all of the data will have to go through that Multiple Communication Unit, there can be many operations providing the latest services for the videoconference-based communication. MCU acts as centralized bridge to facilitate the real time communication between various devices from different places as shown in Fig.3.

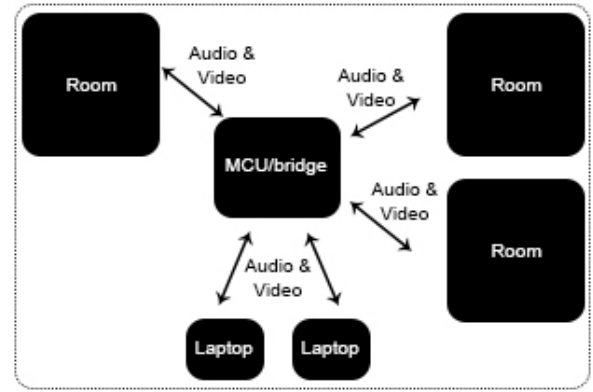

Fig. 3 MCU Bridge for WebRTC

### 4.1.1. Media Transcoding

Using a more cutting-edge Multiple Communication Unit which is capable of mixing and transcoding the streams and flow of media can always pave a way to resolving the heterogeneity of access networks and devices. Through transcoding the streams into diverse sizes and bit-rates, the communication should be possibly modified according to different screen sizes, network situations optimizing the practice of CPU usage and network among the participants leaving it totally to the MCU. This would also help the users in a gateway context where it is required to translate the media streams.

### 4.1.2. Media Composing

Via generating a single audio and video stream from any accessible inputs, the Multiple Communication Unit reduces the volume of CPU overhead and whatever controls that is needed in order to take part in a multi-participant videoconferencing whenever needed.

### 4.1.3. Media Recording

MCU receives whatever streams that are present in the communication. As mentioned before, it is capable to make a composed stream by combining the said streams. In case, the recording of the session is required, MCU stores the flow for any reproductions in future [15]. These capacities thus allow us to offer advanced services that are very often demanded in a multi-conference communication and all collaborative applications by taking control of complete data and information accessible in one session.

### 4.2. WebRTC Multiple Communication Unit Architecture

This part will offer an overview of the structure and architecture of the suggested MCU, which are able to interact with present WebRTC applications. Theoretically, MCU bears four big parts: stream processing, control, transport and API [24]:

#### 4.2.1. API

The functionality is exported by the API. It has to bring up the implementations and functions to specifically point out the streams which are to be received in future, the processes which are to be applied to them, and lastly the output streams. This level also contains the abstract signaling in WebRTC that is capable of establishing the media communications with the participating people.

#### 4.2.2. Transport

In this module, we implement all the functions related to transport. When it comes to the compatibility with WebRTC, transport is the most important part. It brings up the application of ICE along with STUN and TURN as well as RTP and SRTP. Simple WebRTC transmission is shown in Fig.4.

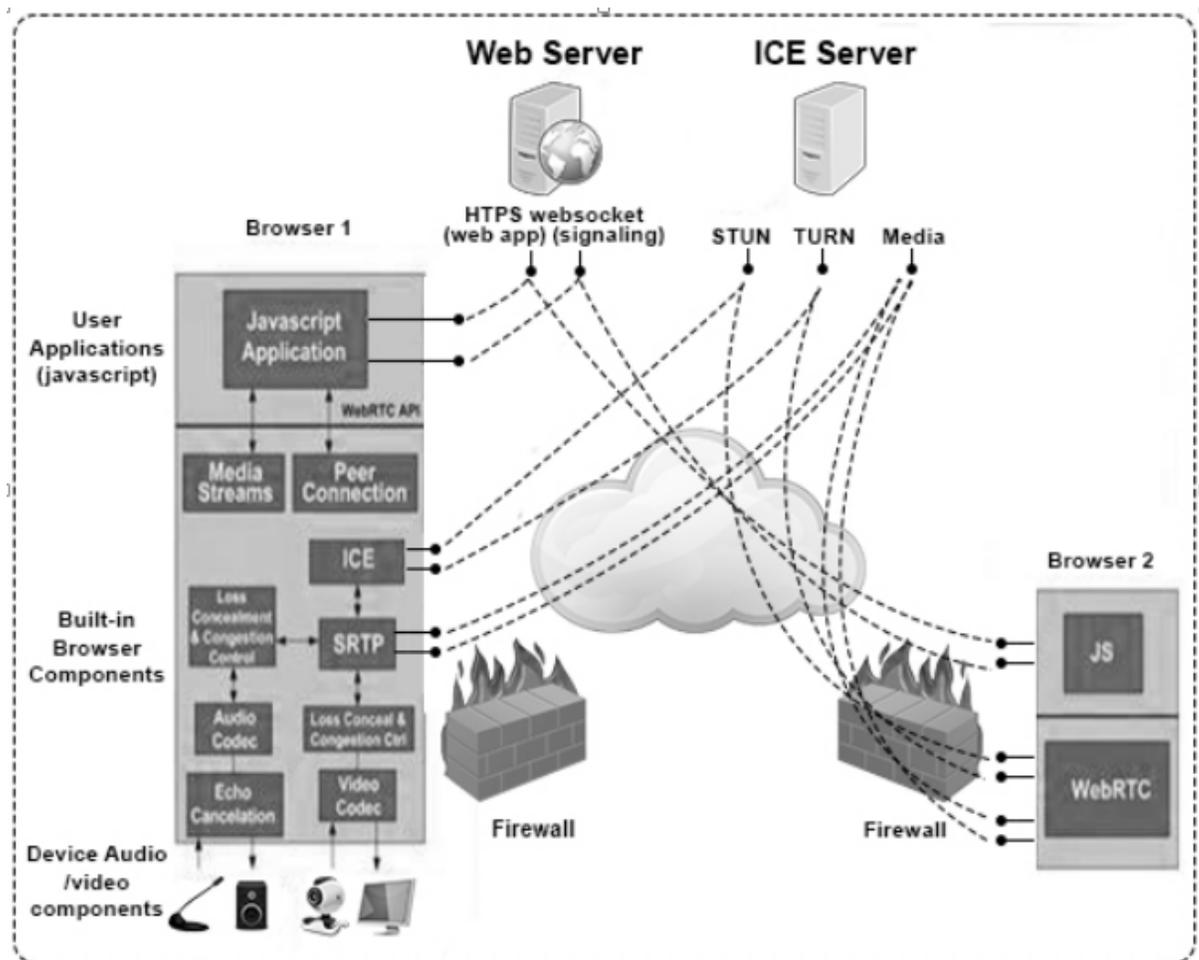

Fig. 4 Simple transmission

### 4.2.3. Stream Processing

This module includes total processing of media by MCU. It can perform processing, mixing, decoding and encoding. It is also able of processing more than one stream and produces more than one output that will eventually be sent through the transport. As it has been revealed previously, the structure of this module accounts to the dividing of process into video and audio processing mechanisms. Each of them bears one or in other cases, more than one decoder, mixer and encoder. Eventually, a video/audio mixer is there only to produce the streams that hold audio as well as video.

### 4.2.4. Control of Stream Processing

This section reflects the completion of the world wide idea of functions or API specified functions. It holds complete control on the transport layers and stream processing. It makes a connection of streams that are received via transport, with appropriately organized stream processors.

### 4.3. Implementation of WebRTC-MCU

After much trial, error, and consideration, we managed to develop a prototype of Multiple Communication Unit application as proposed previously in this paper. In the first stage the emphasis is totally on the component of transport; it is the component that must adhere to the standards strictly. The most crucial stage is the point where it becomes compatible with and according to the present WebRTC implementation. We are trying to use some tested, very familiar and well-documented open-source C libraries e.g. libsrtp and libnice1 [5, 4] for the SRTP and ICE respectively.

However, it comes with its own challenges when one attempt to work with WebRTC at such an early level. The standards and application are both going through a constant evolution and thus an ever-changing environment has come into being. Moreover, a very religious following of the standards is not much possible anymore. The bigger challenge at this level is to improve the communication and adapt constant changes in order to help MCU examine the new alterations or a very well recognized standard. While this paper is being written, our prototype lets the retransmitting of provided WebRTC stream to other members of communication.

## 5. CONCLUSION

In the paper we have proposed a best centralized architecture for the video conferencing to support the WebRTC by using MCU. This is what we call an integrated and centralized architecture of video conferencing which relies on a Multiple Communication Unit for WebRTC ecosystem. We have discussed thoroughly how this structure offers resolutions to certain contexts like stream processing, session recording and composition for screen adaptation as well as bandwidth. A reasonable part of the paper focuses on the challenges that the MCU might face and on the features that are supposed to be fully supported by WebRTC libraries.

The sole purpose behind our attempts and this paper is to show, and in doing so learn, the way WebRTC operates and serves the developers as an adequate starting point for the implementation of other features and for testing the videoconferencing structures. Irrespective of our success in the implementation of MCU, we would like to take its functionality to a whole new level by creating new features like audio/video stream, development to make possible the support for different type of devices (tablets, mobile phones and laptops), gateways for contributing WebRTC users such as XMPP, SIP [23], and also if possible and H.323 [10] clients, streaming and session recording.